\documentclass[pre,twocolumn]{revtex4}
\usepackage{epsfig}
\usepackage{amsmath}
\usepackage{hyperref}
\usepackage{breakurl}

\begin{document}

\title{Chaos and turbulence in clouds}

\author{A. Bershadskii}

\affiliation{
ICAR, P.O. Box 31155, Jerusalem 91000, Israel}

\begin{abstract}

 Spatial buoyancy-helical distributed chaos (turbulence) in the cumulus, stratiform, stratocumulus, cirrus and cirrus mammatus clouds have been studied using results of direct numerical simulations and measurements in cloudy atmosphere. It is shown that in the considered cases the moments of helicity distribution dominate the kinetic energy spectra both in the middle of the clouds and in the cloud-top regions.\\

~~~~~~~~~~~~~~~~~~~~~~~~~~~~~~~~~~~~~~~~~~~~~~~~~~~~~~~~~~~~~~~~~~~Dedicated to the memory of A. Tsinober.\\

\end{abstract}

\maketitle

\section{Inroduction}

  The climate predictions and weather forecasting are heavily dependent on the modelling of the cloud formation under turbulent conditions. Therefore, a vast range of scales should be taken into account - from the cloud microphysical processes to the macroscopic turbulent motions. Both these non-linearly interacting components are poorly understood. The difficulty to conduct measurements of the microphysical properties of the real clouds formation under turbulent conditions (see, for instance, Ref. \cite{si} and references therein) makes the problem formidable. Therefore, the recent direct numerical simulations (see, for instance, Refs. \cite{sg},\cite{li}, \cite{aki} and references therein) trying to take into account the maximum possible variance of the key microscopic and macroscopic processes (collision-coalescence of droplets, condensation-evaporation, the cooling related to the updraft motion inside the clouds, Reynolds number dependent drag, supersaturation, longwave radiative cooling etc.) in the buoyant turbulent environment are rather important and can be compared with results of the numerous atmospheric measurements. \\

  In the Section II of the paper a notion of the distributed chaos has been introduced. In the Section III the buoyancy-helical distributed chaos in the Rayleigh-B\'{e}nard thermal convection has been studied in more detail. In the Section IV the buoyancy-helical distributed chaos in the cumulus clouds has been studied and the consideration has been compared with results a direct numerical simulation. In Sections V-VII the buoyancy-helical distributed chaos has been also studied in the stratiform, stratocumulus, cirrus and cirrus mammatus clouds using results of the direct numerical simulations and the atmospheric measurements in cloudy atmosphere.

\section{From deterministic to distributed chaos}

 The Boussinesq approximation for the buoyancy-driven thermal convection in a layer of fluid (cooled from above and heated from below) in the dimensionless variables is
$$
\frac{\partial {\bf u}}{\partial t} + ({\bf u} \cdot \nabla) {\bf u}  =  -\frac{\nabla p}{\rho_0} + T {\bf e}_z + (Pr/Ra)^{1/2} \nabla^2 {\bf u}   \eqno{(1)}
$$
$$
\frac{\partial T}{\partial t} + ({\bf u} \cdot \nabla) T  =  \frac{1}{(PrRa)^{1/2}} \nabla^2 T, \eqno{(2)}
$$
$$
\nabla \cdot \bf u =  0 \eqno{(3)}
$$
where $p$, ${\bf u}$ and $T$  are the dimensionless pressure, velocity and temperature fields, ${\bf e}_z$ is the vertical unit vector, $Pr$ is the Prandtl number and $Ra$ is the Rayleigh number. \\

  The boundary conditions for the velocity field in the $z$-coordinate are either free-slip or no-slip, and the simplest (Dirichlet) thermal boundary conditions 
$$
T(z=0) = const = T_{bot},~~~~T(z=H) = const = T_{top}   \eqno{(4)}
$$
$H$ is the layer height. In the horizontal $x$ and $y$ coordinate periodic boundary conditions are usually considered. \\
  
   In a recent direct numerical simulation of the Rayleigh-B\'{e}nard convection Eqs. (1-3), reported in Ref. \cite{vss}, power spectra were computed for the vertical component of the velocity field. 
   
   Figure 1 shows such time- and azimuthally averaged spectrum for $Pr =1$, $Ra = 3.85 \times 10^4$, the free-slip and Dirichlet boundary conditions, and  the aspect ratio $\Gamma = L/H = 15$ (where $L$ is the horizontal length of the computational domain). The spectrum was computed for the midplane of the layer. The spectral data were taken from Fig. 9 of the Ref. \cite{vss}.\\ 
   
   The local peak in the Fig. 1 corresponds to the typical horizontal scale of the convection cells (rolls)  $l \ll L$, which form a cell pattern (the `superstructure' in the terms of the Ref. \cite{vss}). \\
   
   The dashed curve in the Fig. 1 indicates the exponential spectrum
$$ 
E(k) \propto \exp(-k/k_c)  \eqno{(5)}
$$ 
\begin{figure} \vspace{-1.5cm}\centering
\epsfig{width=.45\textwidth,file=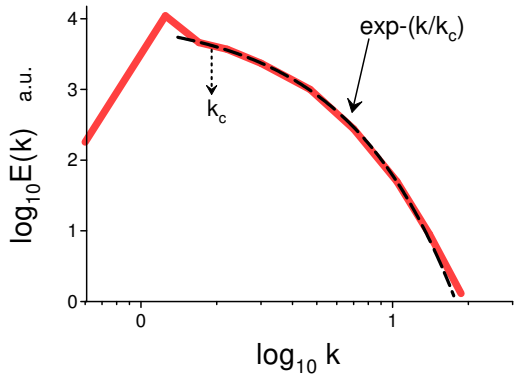} \vspace{-3.8cm}
\caption{Time- and azimuthally averaged spectrum of the vertical velocity in the midplane.}
\end{figure}

   Position of the characteristic scale $k_c$ is shown in the Fig. 1 by a dotted arrow. Such spectrum is a typical characteristic of smooth bounded deterministic dynamics - deterministic chaos (see, for instance, Refs. \cite{fm},\cite{mm} and references therein). \\
   
   Increase of the  Rayleigh number $Ra$ will result in fluctuations of the parameter $k_c$. This can be taken into account with an ensemble averaging
$$
E(k) \propto \int_0^{\infty} P(k_c) \exp -(k/k_c)dk_c \propto \exp-(k/k_{\beta})^{\beta} \eqno{(6)}
$$  
The stretched exponential (in the right-hand side of the Eq. (6)) is a generalization of the exponential spectrum Eq. (5), which preserves the smoothness of the dynamics - distributed chaos. For the distributed chaos the probability distribution $P(k_c)$ at large $k_c$ can be estimated  from Eq. (6) \cite{jon}
$$
P(k_c) \propto k_c^{-1 + \beta/[2(1-\beta)]}~\exp(-\gamma k_c^{\beta/(1-\beta)}) \eqno{(7)}
$$     
(where $\gamma$ is a constant).\\

\section{Helical distributed chaos}

 For the inviscid case ($Ra \rightarrow \infty$) the equation for the mean helicity corresponding to the Eqs. (1-3) is
$$
\frac{d\langle h \rangle}{dt}  = 2\langle {\boldsymbol \omega}\cdot {\bf F}  \rangle \eqno{(8)} 
$$ 
where 
$$
{\bf F} = T {\bf e}_z  \eqno{(9)}
$$
here the vorticity is ${\boldsymbol \omega} = \nabla \times {\bf u}$, the helicity density is $h={\bf u}\cdot {\boldsymbol \omega}$,  and $\langle...\rangle$ corresponds to an average over the liquid volume. As it follows from the Eq. (8) the mean helicity cannot be considered as an inviscid invariant for the thermal convection.  Let us consider, therefore, the case when just the large-scale motions will contribute the main part to the correlation $\langle {\boldsymbol \omega}\cdot {\bf F}  \rangle$, but the correlation is rapidly approaching to zero with reducing spatial scales (it is typical for the turbulent flows). Therefore, inspite of the mean helicity is not inviscid invariant the higher moments of the helicity distribution $h={\bf u}\cdot {\boldsymbol \omega}$ can be approximately considered as the inviscid invariants \cite{lt},\cite{mt}.\\

   To show this, let us divide the liquid volume into a pattern of the cells moving with the liquid - $V_i$ (the Lagrangian description) \cite{lt}\cite{mt}. The boundary conditions on their surfaces should be taken as ${\boldsymbol \omega} \cdot {\bf n}=0$. Moment of order $n$ can be then defined as 
 $$
I_n = \lim_{V \rightarrow  \infty} \frac{1}{V} \sum_j H_{j}^n  \eqno{(10)}
$$
with the total helicity in the subvolume $V_i$
$$
H_j = \int_{V_j} h({\bf r},t) ~ d{\bf r}.  \eqno{(11)}
$$
   Due to the rapid reduction of the correlation $\langle {\boldsymbol \omega}\cdot {\bf F} \rangle$ with the scales the `cell' helicities $H_j$ can be still considered as inviscid invariants for the cells $V_i$ with the small enough spatial scales. These cells will provide the main contribution to the $I_n$ with $n \gg 1 $ for sufficiently chaotic flows (cf. \cite{bt}).  Therefore, the  $I_n$ for the sufficiently large $n$ can be still considered as an inviscid quasi-invariant while the total helicity $I_1$ cannot. For sufficiently chaotic flows the value $n=2$ and $n=3$ can be considered as sufficiently large  (where $I_2$ is the Levich-Tsinober invariant of the Euler equation \cite{lt}). For the viscous cases the `high' moments $I_n$ can be still considered as adiabatic invariants in the inertial range of scales. \\

  Chaotic attractors in the phase space correspond to each of the adiabatic invariants $I_n$. Their basins of attraction can be significantly different: than larger $n$ - thinner the basin of attraction (the intermittency). Therefore, the dynamics of the flow is dominated by the adiabatic invariant $I_n$ with the smallest available order $n$. Let us begin our consideration from  $I_3$ for simplicity. \\                 
  
    The dimensional considerations can be used to estimate characteristic velocity $u_c$ for the fluctuating $k_c$ 
 $$
 u_c \propto |I_3|^{1/6} k_c^{1/2}    \eqno{(12)}
 $$
If one assume a Gaussian (with zero mean) distribution for the characteristic velocity $u_c$  \cite{my}, one obtains from the Eq. (12)
$$
P(k_c) \propto k_c^{-1/2} \exp-(k_c/4k_{\beta})  \eqno{(13)}
$$
where parameter $k_{\beta}$ is a constant.\\

\begin{figure} \vspace{-1.23cm}\centering
\epsfig{width=.43\textwidth,file=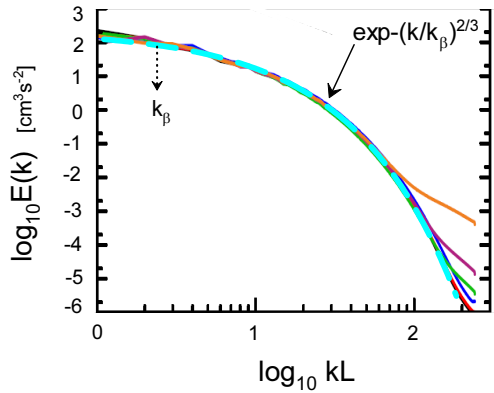} \vspace{-4cm}
\caption{Kinetic energy spectrum for $Re_{\lambda} =104$ (cumulus).} 
\end{figure}
\begin{figure} \vspace{-1cm}\centering
\epsfig{width=.37\textwidth,file=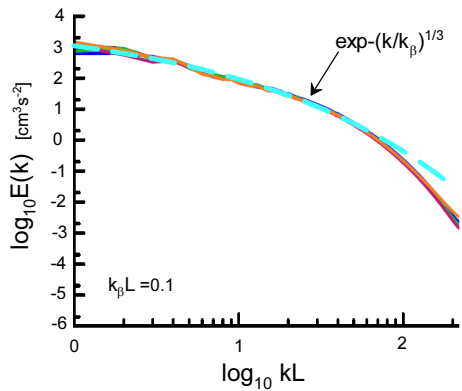} \vspace{-2.7cm}
\caption{As in the Fig 2 but for $Re_{\lambda} =167$. } 
\end{figure}
\begin{figure} \vspace{-0.45cm}\centering
\epsfig{width=.45\textwidth,file=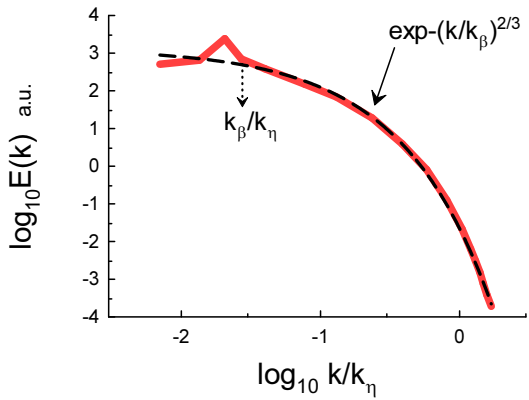} \vspace{-4.2cm}
\caption{Kinetic energy spectrum for $Re_{\lambda} =130$ (stratiform). } 
\end{figure}

   Substituting the Eq. (13) into the Eq. (6) one obtains
$$
E(k) \propto \exp-(k/k_{\beta})^{1/2}  \eqno{(14)}
$$

   Generally the estimate Eq. (12) can be replaced by estimate
$$
 u_c \propto |I_n|^{1/2n}~ k_c^{\alpha_n}    \eqno{(15)}
 $$    
where 
$$
\alpha_n = 1-\frac{3}{2n}  \eqno{(16)}
$$  

 If $u_c$ has Gaussian (normal) distribution a relationship between the considered exponents $\beta_n$ and $\alpha_n$ can be obtained from the Eqs. (7) and (15)
$$
\beta_n = \frac{2\alpha_n}{1+2\alpha_n}  \eqno{(17)}
$$
 
  Substituting $\alpha_n $  from the Eq. (16) into the Eq. (17) one obtains
 $$
 \beta_n = \frac{2n-3}{3n-3}   \eqno{(18)}  
 $$
 For $n \gg 1$ the Eqs. (18) provides
$$
E(k) \propto \exp-(k/k_{\beta})^{2/3}  \eqno{(19)}
$$ 
and for $n=2$ (the Levich-Tsinober invariant)
$$
E(k) \propto \exp-(k/k_{\beta})^{1/3}  \eqno{(20)}
$$

\section{Cumulus clouds - direct numerical simulations}

  In cumulus clouds the buoyancy-driven thermal convection is complicated by collision-coalescence of droplets with
hydrodynamic interaction and Reynolds number dependent drag. Nonlinear interaction between turbulent mixing and the condensation-evaporation process (in presence of large numbers of droplets) should also strongly affect the convection and evolution of cloud droplets.  

   There exist many models for description of this complex system. For instance, it was suggested (see recent Ref. \cite{sg} and references therein) to consider a small air parcel (a cubic box with volume $L^3$) ascending inside the core region of a (maritime) cumulus cloud. In the sufficiently small parcel the fluctuating quantities can be considered as statistically homogeneous and the periodic boundary conditions in all three directions can be used. One have to consider corresponding equations describing dynamics of the parcel, flow and droplets. \\
   
   To describe dynamics of the parcel one can use equations
$$
\frac{\langle d{\mathcal H}(t) \rangle}{dt} =   \langle W(t) \rangle~~~ \frac{d\langle W(t) \rangle}{dt}= \langle B(t) \rangle  \eqno{(22,23)}
$$    
where $B(t)$ is a buoyancy force acting on the parcel, $W(t)$ and $\mathcal H(t)$ are corresponding updraft (vertical) velocity and altitude associated with the parcel.\\
\begin{figure} \vspace{-1.5cm}\centering
\epsfig{width=.45\textwidth,file=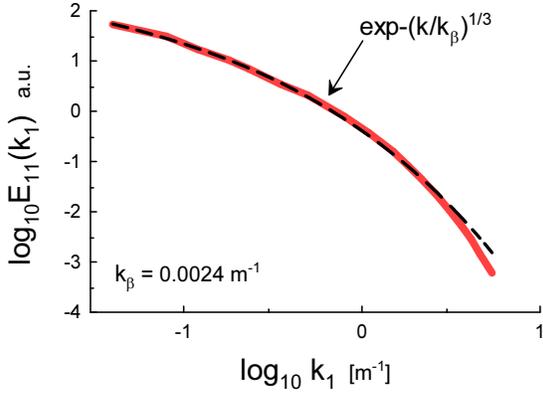} \vspace{-4.3cm}
\caption{Longitudinal spectrum of the horizontal velocity at the middle of the cloud. } 
\end{figure}
\begin{figure} \vspace{-0.5cm}\centering
\epsfig{width=.45\textwidth,file=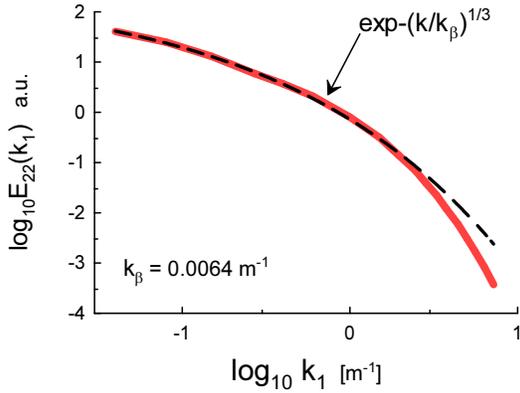} \vspace{-4.4cm}
\caption{Transverse spectrum of the horizontal velocity at the middle of the cloud. } 
\end{figure}
\begin{figure} \vspace{-1.45cm}\centering
\epsfig{width=.45\textwidth,file=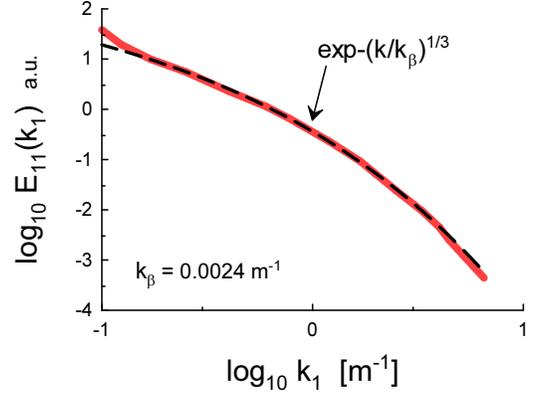} \vspace{-4.4cm}
\caption{Longitudinal spectrum of the horizontal velocity for upper part of cloud.} 
\end{figure}
   
   To describe the flow velocity field ${\bf u}$, temperature $\theta$ and water vapor mixing ratio $q$ one can use the Boussinesq approximation
$$
\frac{\partial {\bf u}}{\partial t} + ({\bf u} \cdot \nabla) {\bf u}  =  -\frac{\nabla p}{\rho_0} + (B -\langle B(t) \rangle ) {\bf e}_z + \nu \nabla^2 {\bf u}  +{\bf f} \eqno{(24)}
$$
$$
\frac{\partial \theta}{\partial t} + ({\bf u} \cdot \nabla) \theta  =  -\Gamma u_z + \frac{L_v}{c_p}(C_d-\langle C_d\rangle)+\kappa \nabla^2 \theta, \eqno{(25)}
$$
$$
\frac{\partial q}{\partial t} + ({\bf u} \cdot \nabla) q  =  -(C_d-\langle C_d\rangle)+\kappa_q \nabla^2 q, \eqno{(26)}
$$
$$
\nabla \cdot \bf u =  0 \eqno{(27)}
$$
in the local coordinate system associated with the parcel. In these equations ${\bf f}$ is an external (random) force, $\Gamma$ represents the cooling effect related to the ascending motion of the parcel, $C_d$ represents latent mass exchange through condensation in the Eq. (26) and heat release in the Eq. (25) (see for more details the Ref. \cite{sg}). \\

   The equation of the mean helicity for the ideal case ($\nu =0$) is
$$
\frac{d\langle h \rangle}{dt}  = 2 \langle {\boldsymbol \omega} \cdot {\bf F} \rangle \eqno{(28)} 
$$
$$
{\bf F} = (B-\langle B \rangle) {\bf e}_z +{\bf f}   \eqno{(29)}
$$   
  If the correlation $\langle {\boldsymbol \omega} \cdot {\bf F} \rangle$ is negligible or it is not negligible for the large scales only then the consideration of the previous Section can be readily generalized for the Eq. (28).\\

   A direct numerical simulation with the above described model was performed in the Ref. \cite{sg} and figures 2 and 3 show the kinetic energy spectrum obtained in this simulation for the Taylor-Reynolds number \cite{my} $Re_{\lambda} = 104$ and $Re_{\lambda} = 167$ (the specral data were taken from Fig. 7 of the Ref. \cite{sg}). Different colors correspond to different times of the system's evolution: from $t=10s$ (black) to $t= 600s$ (orange). The dashed curves indicate correspondence to the Eqs. (19) and (20), correspondingly.

\section{Stratiform and stratocumulus clouds - direct numerical simulations}

   In the stratiform clouds the updraft velocity of the parcel is about zero \cite{hs},\cite{kor}. This results, in particular, in absence of the related cooling effect. Therefore, the condensational growth of the cloud droplets is mainly driven by supersaturation fluctuations (see, for instance, a recent Ref. \cite{li} and references therein).\\

   In the Ref. \cite{li} a direct numerical simulation of a model corresponding to the stratiform clouds was performed with especial attention to the phenomenon of the cloud-droplet growth due to supersaturation fluctuations affected by turbulence. Figure 4 shows a time-averaged kinetic energy spectrum obtained in this simulation for $Re_{\lambda} =130$ (the spectral data were taken from Fig. 1a of the Ref. \cite{li}). The wavenumber is normalized by the Kolmogorov wavenumber $k_{\eta} =2\pi/\eta$, where $\eta = (\langle \varepsilon \rangle /\nu^3)^{-1/4}$ is the Kolmogorov (dissipative) scale and  $\langle \varepsilon \rangle$ is the mean dissipation rate due to viscosity \cite{my}. The dashed curve indicates correspondence to the Eq. (19). \\
      
   In the Ref. \cite{aki} results of a direct numerical simulation of stratocumulus cloud-top turbulence were reported (see  for more details of the setting and simulations in the Refs. \cite{mel},\cite{sm}). In the cloud-top region longwave radiative cooling results in convective instability, that is a major source of chaotic (turbulent) motion in the cloud's region.
   
   Figure 5 shows longitudinal spectrum of the horizontal velocity obtained in this simulation at the middle of the cloud (the spectral data were taken from Fig. 4 of the Ref. \cite{aki}). Figure 6 shows corresponding transverse spectrum. Figure 7 shows longitudinal spectrum of the horizontal velocity obtained in this simulation for the upper part of cloud. The dashed curves indicate the helically dominated distributed chaos Eq. (20).\\
   
\begin{figure} \vspace{-1.35cm}\centering
\epsfig{width=.45\textwidth,file=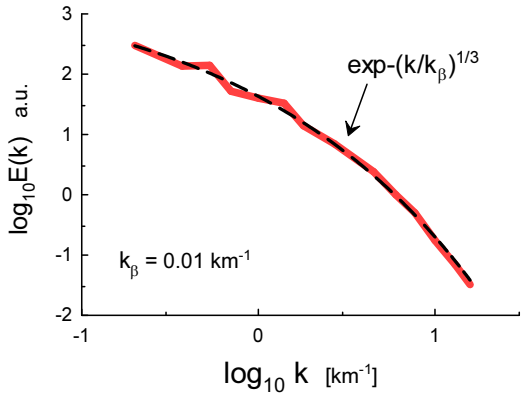} \vspace{-4.45cm}
\caption{1D power spectrum of the radiance in the North-South direction (DF). } 
\end{figure}
\begin{figure} \vspace{-1.43cm}\centering
\epsfig{width=.45\textwidth,file=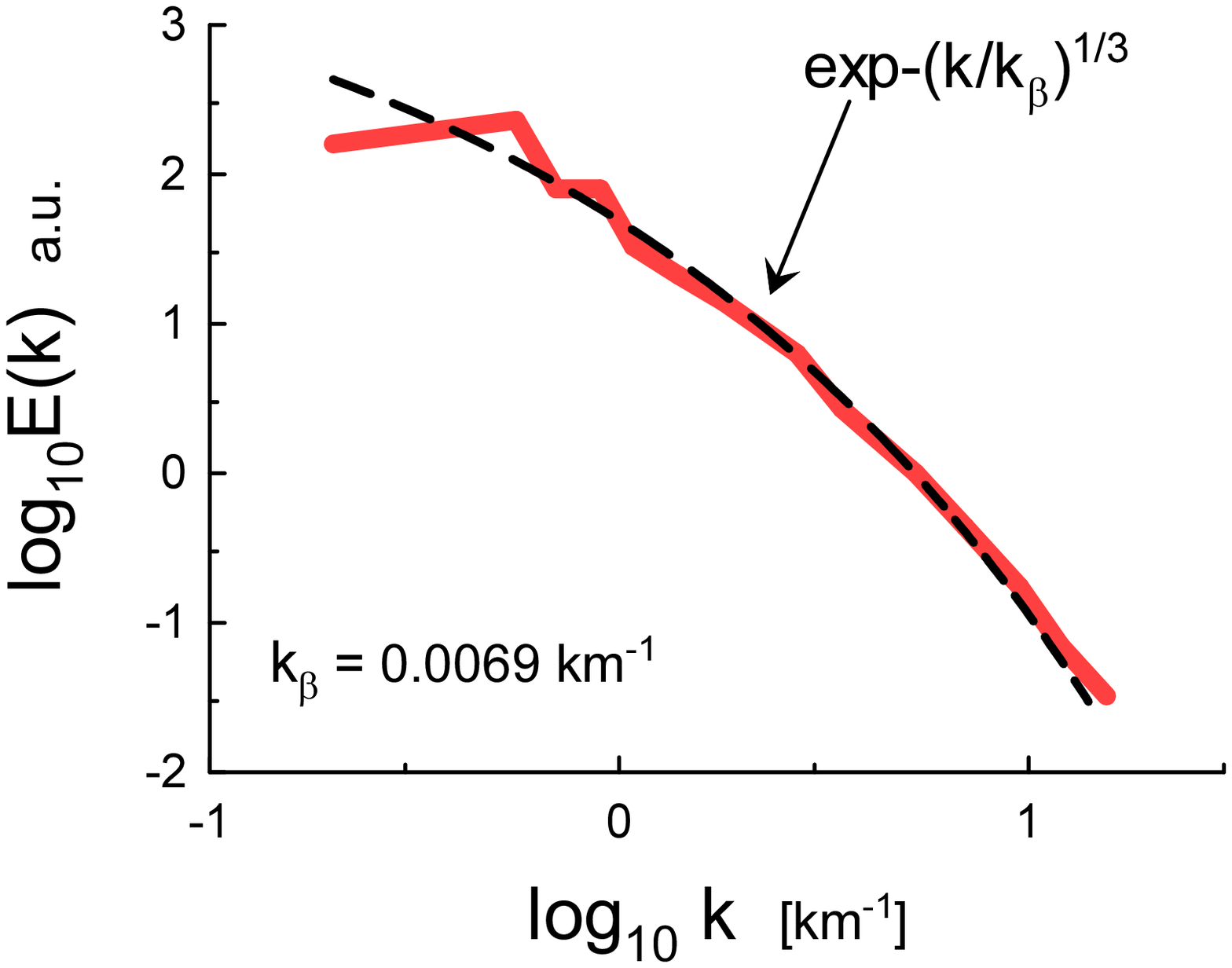} \vspace{-4.4cm}
\caption{1D power spectrum of the radiance in the North-South direction (DA).} 
\end{figure}
\begin{figure} \vspace{-0.5cm}\centering
\epsfig{width=.45\textwidth,file=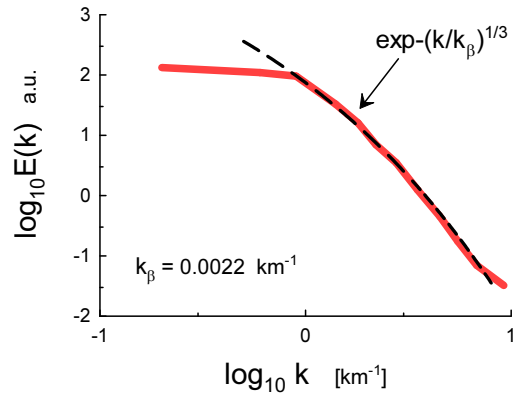} \vspace{-4.45cm}
\caption{1D power spectrum of the radiance in the East-West direction (DF). } 
\end{figure}

\section{Stratocumulus clouds - radiometric and velocity measurements}

\subsection{Over land}   The results of radiometric measurements made by the airborne Multiangle Imaging Spectro-Radiometer (AirMISR), on the NASA ER-2 high-altitude ($\sim 20$ km) aircraft, were reported in Ref. \cite{om}. The measurements were made for the blue channel radiance ($\lambda = 0.443~ \mu$m) over continental single-layer low stratocumulus clouds. Therefore, one can expect that the radiance fluctuations were produced mainly by the velocity field at the top of the clouds. 

  Figure 8 shows 1D power spectrum of the radiance in the North-South direction: zenith angle $70.5^o$ and scattering angle $140.8^o$ (the spectral data were taken from Fig. 7-DF of the Ref. \cite{om}).  Figure 9 shows 1D spectrum of the radiance in the North-South direction: zenith angle $70.5^o$ and scattering angle $72.3^o$ (the spectral data were taken from Fig. 7-DA of the Ref. \cite{om}). The DF and DA camera's view angles are the most oblique ones and for them the cloud gaps close about completely.\\

   Figure 10 shows 1D spectrum of the radiance in the East–West direction (the spectral data were taken from Fig. 8-DF of the Ref. \cite{om}).  Figure 11 shows 1D spectrum of the radiance in the the East-West direction direction (the spectral data were taken from Fig. 8-DA of the Ref. \cite{om}). The dashed curves in the Figs. 8-11 indicate the helically dominated distributed chaos Eq. (20).
   
 \subsection{Over ocean}
 
\begin{figure} \vspace{-1.7cm}\centering
\epsfig{width=.45\textwidth,file=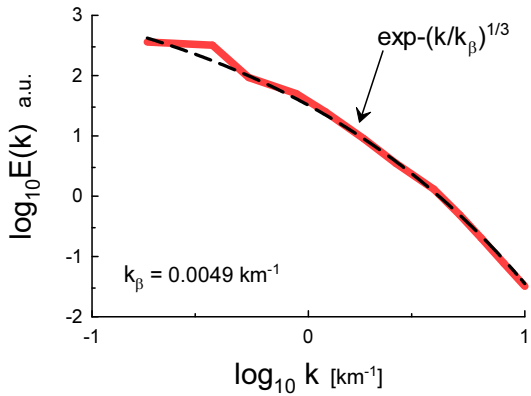} \vspace{-4.25cm}
\caption{1D power spectrum of the radiance in the East-West direction (DA). } 
\end{figure}
  
\begin{figure} \vspace{-0.5cm}\centering
\epsfig{width=.45\textwidth,file=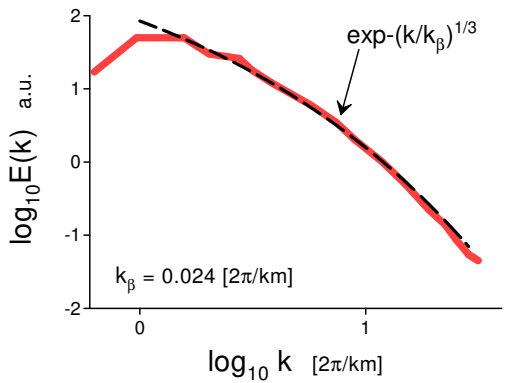} \vspace{-3.8cm}
\caption{Power spectrum of the cloud top reflectance over the marine stratocumulus clouds. } 
\end{figure}
\begin{figure} \vspace{-1.5cm}\centering
\epsfig{width=.47\textwidth,file=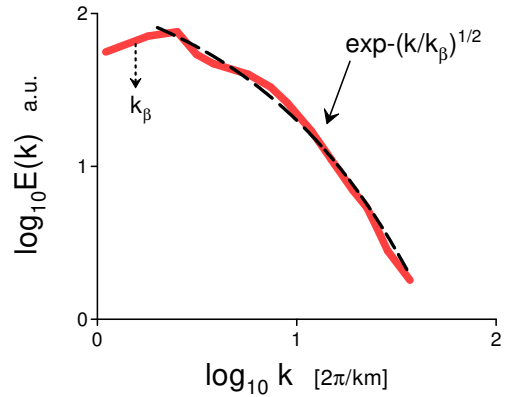} \vspace{-3.9cm}
\caption{Power spectrum of the vertical velocity based on the in-cloud horizontal flights.} 
\end{figure}

    The results of radiometric measurements made by the airborne Multispectral Cloud Radiometr (MCR), on the NASA ER-2 high-altitude ($\sim 20$ km) aircraft, were reported in Ref. \cite{bsh}. The measurements were made for the visible channel radiance ($\lambda = 0.754~ \mu$m) over the marine stratocumulus clouds (scanning was made in the direction perpendicular to the line of flight).\\ 
    
    Figure 12 shows the spectrum of the cloud top reflectance. The spectral data were taken from Fig. 9 of the Ref. \cite{bsh}. The dashed curve in the Fig. 12 indicates the helical distributed chaos Eq. (20).\\
    
    Figure 13 shows power spectrum of the vertical velocity based on the direct measurements of the velocity during the in-cloud horizontal flight (the spectral data were taken from Fig. 8b of the Ref. \cite{bsh}). The dashed curve in the Fig. 13 indicates the helical distributed chaos Eq. (14).
    
  \section{Cirrus and cirrus mammatus clouds}
  
\begin{figure} \vspace{-1.2cm}\centering
\epsfig{width=.45\textwidth,file=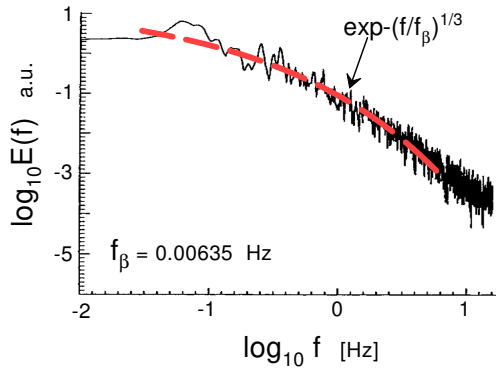} \vspace{-4.3cm}
\caption{Power spectrum of the vertical velocity fluctuations (cirrus clouds).} 
\end{figure}
\begin{figure} \vspace{-0.5cm}\centering
\epsfig{width=.45\textwidth,file=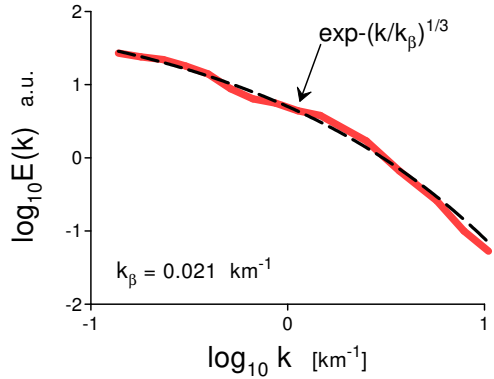} \vspace{-3.7cm}
\caption{Power spectrum of the Doppler velocity (cirrus mammatus clouds).} 
\end{figure}

  Cirrus clouds are composed mainly of ice and are covering extensive areas in the upper troposphere. Therefore, they play the important
role in the earth's radiation budget. The cirrus cloud structure and their optical properties significantly depend on the inner turbulent processes.\\

  In the paper Ref. \cite{sj} results of aircraft measurements made in relatively thick frontal cirrus clouds during the daytime were reported.\\
  
     Figure 14 shows power spectrum of the vertical velocity fluctuations measured at the straight horizontal flights of the aircraft over the region of Wick in
Scotland at a height of 8.9 km. This is a frequency spectrum (the spectral data were taken from Fig. 7a of the Ref. \cite{sj}). However, the frequency spectrum can be readily converted into the wavenumber spectrum using the Taylor hypothesis \cite{my} with $f = U_0~k/2\pi$, where $k$ is horizontal wavenumber and $U_0$ is the aircraft velocity. The dashed curve indicates the helically dominated distributed chaos Eq. (20).\\

   Mammata are protuberances on the undersides of the clouds. The observations show \cite{ws} that cirrus mammata belong to the transition zone from dry air to the cloudy (moist) layers. In the paper Ref. \cite{ws} the Doppler velocity data inferred from the cirrus cloud-base region (a 10-yr subset of high-cloud radar data)  were used to compute the power spectra of the Doppler velocity in the cirrus mammatus clouds.\\
   
    Figure 15 shows power spectrum corresponding to the altitude 7.9 km (the spectral data were taken from the Fig. 7a of the Ref. \cite{ws}). The dashed curve indicates the helically dominated distributed chaos Eq. (20).\\

\section{Acknowledgement}

I thank E. Levich for stimulating discussions, and A.S. Pikovsky and J. Schumacher for consultations.

\end{document}